\documentstyle[aps,12pt,tighten]{revtex}
\begin{document}
\draft
%\preprint{}
\input{epsf}

\title{ Interpreted History Of Neutrino Theory Of Light And Its Future}

\author{W. A. Perkins}
%\address{(Perkins Advanced Computer Systems) Auburn, California }

\address{Perkins Advanced Computing Systems,\\ 12303 Hidden Meadows
Circle, Auburn, CA 95603, USA\\E-mail: wperkins@aub.com} 

\maketitle

\begin{abstract}
De Broglie's original idea that a photon is composed of a 
neutrino-antineutrino pair bound by some interaction 
was severely modified by Jordan. Although Jordan addressed 
an important problem (photon statistics) that de Broglie 
had not considered, his modifications may have been detrimental 
to the development of a composite photon theory. 
His obsession with obtaining Bose statistics for the 
composite photon 
made it easy for Pryce to prove his theory untenable. 
Pryce also indicated that forming transversely-polarized photons 
from neutrino-antineutrino pairs was impossible, 
but others have shown that this is not a problem. 
Following Pryce, Berezinskii has proven that any composite 
photon theory (using fermions) is impossible if one accepts 
five assumptions. Thus, any successful composite theory must show 
which of Berezinskii's assumptions is not valid. 
A method of forming composite particles based on Fermi and Yang's 
model will be discussed. Such composite particles have properties 
similar to conventional photons. However, unlike the situation 
with conventional photon theory, Lorentz invariance can be 
satisfied without the need for gauge invariance or introducing 
non-physical photons.
\end{abstract}


\section{Introduction}
\label{sec.intro}

The theoretical development of a composite photon theory 
(based on de Broglie's idea that the photon is composed of a 
neutrino-antineutrino pair~\cite{broglie} bound by some 
interaction) has had a stormy history 
with many negative papers. 
While de Broglie did not address
the problem of statistics for the composite photon,  
``Jordan considered the essential part of the problem 
was to construct 
Bose-Einstein amplitudes from Fermi-Dirac amplitudes.'' 
Furthermore, ``he suggested that it is not the interaction 
between neutrinos and antineutrinos that binds them 
together into photons, but rather the manner in which 
they interact with charged particles that leads 
to the simplified description of light in terms of photons.''

Jordan's hypothesis~\cite{jordan1935} traded 
one problem for another. 
Although he eliminated the need for theorizing 
an unknown interaction, his hypothesis that the 
neutrino and antineutrino are emitted in exactly 
the same direction seems rather artificial and was 
criticized by Fock~\cite{fock}. His strong desire to obtain
exact Bose commutation 
relations for the composite photon led him to work 
with a scalar or longitudinally polarized photon. 
Indeed, starting with Jordan's $ \ll ansatz \gg $,
\begin{equation}
 {\cal  A}_{\mu} (k)
=\int_0^1 f (\lambda) \bar \psi ((\lambda -1) k) \gamma_{\mu} \psi
(\lambda k) d\lambda,
\label{ansatz}
\end{equation}
Barbour {\it et. al.}~\cite{barbour} 
showed that the resulting photon would be 
longitudinally polarized instead of transversely polarized 
like real photons. If Jordan had worked with a 
transversely polarized photon, he could not have 
obtained Bose commutation relations.

More recently, researchers~\cite{sahlin,penney,lipkin,perkins1972} 
have been satisfied 
that composite particles formed of fermion pairs (quasi-bosons) 
such as the deuteron and Cooper pairs are 
only approximate bosons as they do not quite satisfy 
Bose commutation relations.

Quasi-bosons obey the commutation relations of the form,
\begin{eqnarray}
\left[Q({\bf k}),Q({\bf l})\right] = 0,  \nonumber \\
\left[Q^\dagger({\bf k}),Q^\dagger({\bf l})\right] = 0,\nonumber \\
\left[Q({\bf k}),Q^\dagger({\bf l})\right] 
= \delta({\bf k}-{\bf l})- \Delta({\bf k},{\bf l}).
\label{eqnquasi}
\end{eqnarray}
These are identical to Bose commutation relations except for the
additional $\Delta( {\bf k}, {\bf l} )$ term 
(see Ref.~\cite{lipkin}), whose value is very small. 
Thus, it is easy 
to envisage that Eq.~(\ref{eqnquasi}) is a more accurate form of
the commutation relations for integral spin particles.
 
As presented in many quantum mechanics texts 
it may appear that Bose 
commutation relations  follow from basic principles, 
but it is really from 
the classical canonical formalism. This is {\it not}
 a reliable procedure
as evidenced by the fact that it gives the completely 
wrong result for spin 1/2 particles. Furthermore,
in extending the classical
canonical formalism for the photon, it is necessary to deviate
from the canonical rules (see Ref.~\cite{bjorken_qf}, p.~98).

If Jordan had taken the approach that the composite 
photon is only an approximate boson, Pryce~\cite{pryce} 
could never have
proven that the theory was impossible. Jordan's hypothesis 
that a single neutrino or antineutrino 
can simulate a photon, which was needed to obtain Bose statistics,
is not in agreement with experiment. 
For otherwise, the interaction cross-section for neutrinos 
with matter would be orders of magnitude greater than 
the measured value.

From reading Pryce's paper one might get the impression 
that creating transversely-polarized photons from neutrinos 
is a problem for the theory. The real problem (as Pryce does shows) 
is in obtaining both Bose statistics and transversely-polarized 
photons.  It is interesting 
to note in retrospect that Pryce's four postulates 
do not include the key 
assumption: {\it the photon
obeys Bose commutation relations}, which is essential 
to the theorem. This is
not too surprising since Jordan~\cite{jordan1935} 
and those working with him on the
theory also took Bose statistics as an absolute. 
Berezinskii~\cite{berez} in reaffirming Pryce's theorem, does
include Bose statistics as one of his five assumptions. He argues
that certain Bose commutation relations are necessary for the 
photon to be truly neutral. However, Perkins~\cite{perkins1972}
has shown that a neutral photon in the usual sense can be 
obtained without Bose commutation relations.

Pryce made the difficulties look much worse 
than they are by accepting Jordan's solution to the statistics
problem which sealed the impossibility of solving the polarization
problem. Case~\cite{case} and
Berezinskii~\cite{berez} agree that constructing transversely 
polarized photons is not the problem. Kronig~\cite{kronig} and 
Perkins~\cite{perkins1965,perkins1972} 
have explicitly constructed transversely-polarized photons 
from neutrinos.

In a remarkable paper~\cite{kronig} 
Kronig showed a second quantization 
method for obtaining the photon field from components of the 
neutrino and antineutrino fields. The photons thus obtained 
have transverse polarization and obey Maxwell's equations. 
(Pryce's criticisms do not apply to these calculations.) 
In order to obtain the 
usual commutation relations for the electromagnetic field, 
he introduced a relation between the neutrino spinors A and C 
which is not invariant under rotations of the coordinate system. 
Although Pryce emphasized this error in Kronig's paper, 
the results that Kronig was trying to obtain follow directly 
from plane-wave spinors as shown by Perkins~\cite{perkins1965}.

Although conventional photon theory has been very successful, 
it has some shortcomings. To satisfy Lorentz invariance, it 
is necessary to introduce
non-physical polarization states. 
The difficulty in conventional theory of quantizing 
the electromagnetic field
has been noted~\cite{bjorken_qf}, ``It is ironic that of the fields 
we shall consider it is the
most difficult to quantize.'' To avoid the problems of 
non-physical photons,
Veltman~\cite{veltman} quantized the field with a small photon 
mass. More serious are the divergent 
problems which require renormalization. A composite photon 
might help in solving some of these problems.

Two major problems facing the theory are: (1) Understanding 
the neutrino-antineutrino interaction that ``binds'' them into 
a photon, and (2) showing
that the non-Bose photon (which results from the theory) is not 
in contradiction with any physical laws.
 
This latter point 
is the subject of a recent paper~\cite{perkins1999} in which it 
is shown that the Blackbody radiation spectrum for composite 
photons is so similar to Planck's law that existing experiments
could not have detected the difference. 
The commutation 
relations for the fields do
not satisfy space-like commutativity, 
but this is true of many integral spin
particles and just indicates that composite 
particles have a finite extent~\cite{perkins1999}.

According to the Standard Model, the neutrino is 
described by a two- component theory. 
To form a photon which satisfies parity and charge
conjugation, we need both sets of two-component 
neutrinos, right-handed
and left-handed neutrinos. Two sets of neutrinos 
have been observed, one
that couples with electrons and one that couples 
with muons. These can be
our two sets of two-component neutrinos if the 
positive muon is identified as
the particle and the negative muon as the 
antiparticle (see Sec. IV of Ref.~\cite{perkins1965}). 

\section{Forming Composite Particle Wave Functions}
\label{sec.forming} 

Fermi and Yang~\cite{fermi-yang} have outlined 
a method of forming a 
composite pion
from a nucleon-antinucleon pair. They imposed some 
special requirements
on the interaction. The attraction should not be a 
new force field, for this
would require the quanta of that new field to be 
elementary particles
themselves. Thus, only forces of zero range are 
compatible with relativistic
invariance and the simplest relativistically invariant 
interactions between two
fields are usual five types~\cite{fermi-yang}.
%\begin{eqnarray}
%\int d^3 r \: \psi^\dagger \gamma_4 \psi \; (scalar) \nonumber \\
%\int d^3 r \: \psi^\dagger \gamma_4 \gamma_5 \psi \; 
%(pseudoscalar) \nonumber \\
%\int d^3 r \: \psi^\dagger \gamma_4 \gamma_{\mu} 
%\psi \; (vector) \nonumber \\
%\int d^3 r \: \psi^\dagger \gamma_4 \gamma_{\mu} \gamma_5 
%\psi \; (pseudovector) \nonumber \\
%\int d^3 r \: \psi^\dagger \gamma_4 \sigma_{\mu \nu}\psi \; 
%(tensor) \nonumber \\ 
%\label{eqn1}
%\end{eqnarray}

We want to use these interactions for the same reasons 
that Fermi and
Yang suggested them and in addition because we know 
from our previous
work~\cite{perkins1965,perkins1972} 
that they result in transversely-polarized 
photons. Differing from
Fermi and Yang, we will use two-component neutrinos 
instead of nucleons
and work at the two-component level. We prefer to 
work at the two-component level as all 
terms are meaningful there, 
unlike the 4-component
level where most of the terms are zero.

The Weyl equations for the two types of neutrinos are: 
\begin{eqnarray}
{\bf \sigma} \cdot {\bf p} \; \phi_{\nu 1}(x) 
= i {\partial \phi_{\nu 1}(x) \over \partial t}, \nonumber \\
{\bf \sigma} \cdot {\bf p} \; \phi_{\nu 2}(x) 
= -i {\partial \phi_{\nu 2}(x) \over \partial t}, 
\label{eqn2}
\end{eqnarray}
where the $\sigma$'s are the Pauli 2x2 spin matrices. 
These can be put into a
symmetric (covariant) form (similar to the 4-component 
$\gamma_\mu p_\mu \Psi(x)= 0$ ) by
defining $\sigma^1_\mu = (\sigma,iI)$ and 
$\sigma^2_\mu = (\sigma,-iI)$ with $I$ being the unit matrix. The
equations in (\ref{eqn2}) become,
\begin{eqnarray}
\sigma^1_\mu  p_\mu \phi_{\nu 1}(x) = 0, \nonumber \\
\sigma^2_\mu  p_\mu \phi_{\nu 2}(x) = 0. 
\label{eqn3}
\end{eqnarray}
The spinors obtained from plane-wave solutions 
for $\nu_1$ and $\nu_2$ respectively
(positive energy) are:
%\phi_{\nu 1}(x) = u({\bf p}) e^{i p x} \nonumber \\
%\phi_{\nu 2}(x) = v({\bf p}) e^{i p x} 
%\label{eqn4}
%\begin{eqnarray}
%\end{eqnarray}
\begin{eqnarray}
u({\bf p}) = \sqrt{ {p_4 + i p_3} \over 2 p_4} 
\left( \begin{array}{c}
 1 \\ {{p_1 + i p_2} \over {p_3 - i p_4}} 
\end{array} \right), 
\nonumber \\
v({\bf p}) = \sqrt{ {p_4 + i p_3} \over 2 p_4}
\left(  \begin{array}{c}
{{-p_1 + i p_2} \over {p_3 - i p_4}} \\ 1 
 \end{array} \right), 
\label{eqn5}
\end{eqnarray}
where $p_\mu$ is $({\bf p},ip)$.
They have been normalized in the sense that 
$\phi^\dagger(x) \phi(x) = 1$. Note that $u({\bf p})$
and $v({\bf p})$ 
depend only upon the direction of the momentum 
(${\bf n} = {\bf p} / |{\bf p} |$),
and not its magnitude. For negative momentum (${\bf -n}$), 
we obtain from (\ref{eqn3}) the relations,
\begin{eqnarray}
u({\bf -p}) = v({\bf p}),  \nonumber \\
v({\bf -p}) = u({\bf p}).  
\label{eqn6}
\end{eqnarray}

Operating on these solutions with the spin operator
shows that $u({\bf p})$ corresponds to spin parallel to direction 
of propagation
while $v({\bf p})$ is antiparallel. To obtain the antineutrino 
wave functions, we
take the negative energy solutions and operate on 
them with CP (see p. 260
of Ref.~\cite{bjorken_qm}),
\begin{equation}
\phi_{CP}({\bf x},t) = C \phi^*({\bf -x},t) 
= \mp i \sigma_2 \phi^*({\bf -x},t). 
\label{eqn8}
\end{equation}
Surprising, this results in each antineutrino spinor 
being identical with its
neutrino spinor! The combined wave function 
for both set of
neutrinos is:
\begin{eqnarray}
\Psi(x) = {1 \over \sqrt{V}} \sum_{\bf k} \left\{ 
\left[ a_1({\bf k})u({\bf k}) +  a_2({\bf k})v({\bf k}) 
\right] e^{i k x} \right. \nonumber \\
\left. + \left[c_1^\dagger({\bf k})u({\bf k}) 
+  c_2^\dagger({\bf k})v({\bf k}) \right]e^{-i k x} \right\},
\label{eqn9}
\end{eqnarray}
where $k x$ stands for $ {\bf k} \cdot {\bf x} + k_4 x_4 =
{\bf k} \cdot {\bf x} - \omega_k t$.
$a_1$ and $c_1$ are the fermion annihilation operators for $\nu_1$
and $\overline \nu_1$
respectively, while $a_2$ and $c_2$  are 
the annihilation operators for $\nu_2$
and $\overline \nu_2$.
%Note that there is no ${1 \over \sqrt{k_0}}$ 
%factor in (\ref{eqn9}). 
%It is effectively 
%contained in the
%spinors, $u({\bf p})$ and  $v({\bf p})$, 
%which are normalized so that 
%$u^\dagger u = v^\dagger v = 1$. This
%normalization will be useful in the formation of 
%composite particles.

We will assume that during an interaction 
(creation or annihilation) with
other particles, the neutrino-antineutrino pair 
have antiparallel momenta.
All possible combinations of such 
neutrino-antineutrino pairs are indicated
in Fig.~1. We take the composite particle field
$G_{int}(R)$ to be the
superposition of all the combinations in Fig.~1 
with $O_{int}$ representing
the neutrino-antineutrino interaction. 
The composite-particle coordinate is ${\bf R}$ 
and its momentum and energy are ${\bf P}$ and $\omega_P$. 
One particle has momentum ${\bf P + k}$
while the other has momentum  ${\bf -k}$
with  ${\bf k}$
parallel to ${\bf P}$.
\begin{eqnarray}
G_{int}(R) =  \sum_{\bf P} {1 \over {2 \sqrt{ V \omega_p}}}
\sum_{\bf k} \left\{ F^\dagger(k, {\bf n}) \right. \nonumber\\ \left.
[ c_1(k,{\bf -n})u^\dagger({\bf -n})O_{int}a_1(P+k,{\bf n})
u({\bf n}) + ... ] e^{i P R}  
\right. \nonumber \\ \left. 
+ F(k, {\bf n}) [ a_1^\dagger(P+k,{\bf n})u^\dagger({\bf n})
O_{int}c_1^\dagger(k,{\bf -n})u({\bf -n}) + ...]e^{-i P R} \right\}.
\label{eqn10}
\end{eqnarray}
The individual terms of Eq.~(\ref{eqn10})
correspond to the neutrino-antineutrino pairs 
in Fig.~1. Only the first of the eight terms is shown in
(\ref{eqn10}) for brevity.
If we tried to
combine {\it massive} particles in this manner, it 
would not work because $u({\bf P + k}) \ne u({\bf k})$ 
in that case. 

\vskip0.25in

\epsfxsize=5.0in \epsfbox{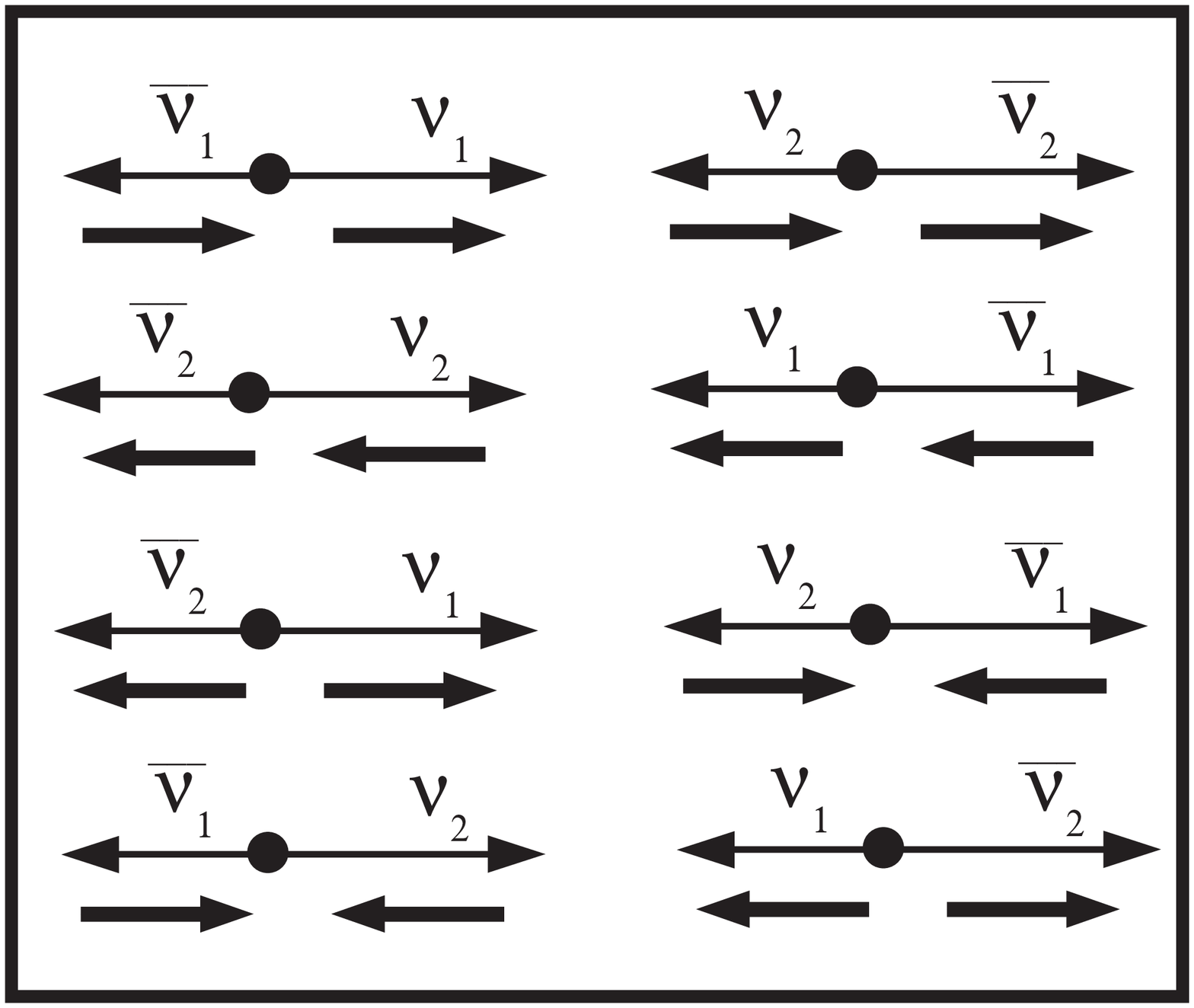}

\begin{figure}[t]
\caption{All neutrino-antineutrino combinations with antiparallel
momenta. The upper arrows indicate the momentum direction with
the longer arrows corresponding to ${\bf P} + {\bf k}$ and the
shorter ones to ${\bf -k}$. The lower, wider arrows indicate spin
direction. Each neutrino-antineutrino combination corresponds
to an annihilation term and a creation term of Eq.~(\ref{eqn10}).}
\label{fig1}
\end{figure}

If we try to identify the field of Eq.~(\ref{eqn10}) 
with the photon, 
there is clearly a problem as some terms correspond to
longitudinal polarization (spin-0). 
It has been noted that one could consider each transverse 
polarization as an
independent particle (see Ref.~\cite{veltman}, p. 173, 180-2). 
This is because the
polarization vectors are independent degrees of 
freedom and under a Lorentz
transformation change into themselves. 
(For invariance under parity one
needs both transverse polarizations.) 
Therefore, it seems logical to consider
that $G_{int}(R)$ is composed of two fields: 
the photon and another particle with
longitudinal polarization.

Using (6), the terms corresponding to spin-1 
(transverse polarization) in Eq.~(10) can be put in the form:
\begin{eqnarray}
A_\mu(R) =  \sum_{\bf P} {1 \over \sqrt{2 V \omega_p}}\left\{ 
\left[\gamma_R({\bf P})\epsilon_\mu^1({\bf n})
+ \gamma_L({\bf P})\epsilon_\mu^2({\bf n}) 
\right]e^{i P R} \right. \nonumber \\
\left. + \left[\gamma_R^\dagger({\bf P})\epsilon_\mu^{1*}({\bf n})
+ \gamma_L^\dagger({\bf P})\epsilon_\mu^{2*}({\bf n}) 
\right]e^{-i P R}  \right\},
\label{eqn13}
\end{eqnarray}
where
\begin{eqnarray}
\gamma_R({\bf P}) =  \sum_{\bf k} F^\dagger( k, {\bf n}) 
\left[ c_1(k, {\bf -n}) a_1(P+k,{\bf n}) 
+  c_2(P + k, {\bf n}) a_2(k,{\bf -n}) \right], \nonumber \\
\gamma_L({\bf P}) =  \sum_{\bf k} F^\dagger( k, {\bf n}) 
\left[ c_2(k, {\bf -n}) a_2(P+k,{\bf n}) 
+  c_1(P + k, {\bf n}) a_1(k,{\bf -n}) \right]. 
\label{eqn14}
\end{eqnarray}
can be identified as composite-particle 
annihilation operators. The spinors combinations in (\ref{eqn10})
are only a function of ${\bf n}$, 
and they are obviously
the polarization vectors of the composite particle. 
Two choices for $O_{int}$ are $\sigma_\mu^1 $
and $\sigma_\mu^2 $, both of which result in combinations that 
transform like four-vectors. For transverse polarization the
result is the same with either one, 
\begin{eqnarray}
\epsilon_\mu^1( n) = {1 \over \sqrt{2}} v^\dagger( {\bf n} )
\sigma_\mu^1 u( {\bf n}), \nonumber \\
\epsilon_\mu^2( n) = {1 \over \sqrt{2}} u^\dagger( {\bf n} )
\sigma_\mu^1 v( {\bf n}).
\label{eqn17}
\end{eqnarray}
Carrying out the matrix multiplications results in,
\begin{eqnarray}     
\epsilon_\mu^1(n) \!= \!{1 \over \sqrt{2}} \left( 
{{-i n_1 n_2 \!+\!1 \!+\!n_3 \!-\!n_1^2} \over {1 + n_3}},
{{- n_1 n_2 \!+ \!in_1^2 \!+ \!in_3^2 \!+ \!in_3} 
\over {1 + n_3}},
\!-n_1 \!- \!i n_2, 0 \right), \nonumber\\ 
\epsilon_\mu^2(n) \!= \!{1 \over \sqrt{2}}\left( 
{{i n_1 n_2 \!+\!1 \!+\!n_3 \!-\!n_1^2} \over {1 + n_3}},
{{- n_1 n_2 \!- \!in_1^2 \!- \!in_3^2 \!- \!in_3} 
\over {1 + n_3}},
\!-n_1 \!+ \!i n_2, 0 \right).  
\label{eqn18}
\end{eqnarray}     
These polarization vectors satisfy the 
normalization relation, $\epsilon_\mu^j(n) \cdot 
\epsilon_\mu^{j*}(n) = 1$.
The four-vectors, $\epsilon_\mu^1(n)$  and $\epsilon_\mu^2(n)$ 
are orthogonal as $\epsilon_\mu^1(n) \cdot 
\epsilon_\mu^{2*}(n) = 0$.
The Lorentz-invariant dot 
products of the momentum $P_\mu = |{\bf P}|(n_1,n_2,n_3,i)$
with the polarization vectors are,
\begin{eqnarray}     
P_\mu \epsilon_\mu^1(n) = 0, \;
P_\mu \epsilon_\mu^2(n) = 0,
\label{eqn26}
\end{eqnarray}     
and in three dimensions, 
\begin{eqnarray}     
{\bf n} \cdot  {\bf \epsilon^1}({\bf n}) = 
{\bf n} \cdot  {\bf \epsilon^2}({\bf n}) = 0, \;
{\bf \epsilon^1}({\bf n}) \times 
{\bf \epsilon^2}({\bf n})= -i{\bf n}, 
 \nonumber\\
{\bf n} \times {\bf \epsilon^1}({\bf n})=-i 
{\bf \epsilon^1}({\bf n}), \;
{\bf n} \times {\bf \epsilon^2}({\bf n})=i 
{\bf \epsilon^2}({\bf n}).
\label{eqn27}
\end{eqnarray}     
If the momentum 
is along the third axis, the
polarization vectors reduce to,
\begin{eqnarray}     
\epsilon_\mu^1(n) = {1 \over \sqrt{2}}(1,i,0,0), \nonumber\\ 
\epsilon_\mu^2(n) = {1 \over \sqrt{2}}(1,-i,0,0),  
\label{eqn28}
\end{eqnarray}     
which are the usual polarization vectors 
for right and left
circular-polarized photons respectively.

\section{The Field of the Composite Photon}
\label{sec.field}
                                    
In this section, we will compare the 
similarities and
differences between the composite photon field and the 
conventional theory. 

\subsection{Similarities to Conventional Theory}
\label{sec.similar}

In comparing the similarities, it should be noted that  
 $\epsilon_\mu^1(n)$ and $\epsilon_\mu^2(n)$ are
the same as the usual right and left circular 
polarization vectors for ${\bf n}$ along
the third axis. However, an advantage of the 
composite theory is that Eq.~(\ref{eqn18}) 
gives the polarization vectors for any ${\bf n}$.

From $A_\mu(R)$, we can readily obtain the Hermitian 
fields ${\bf E}$ and ${\bf H}$.
They follows directly from 
differentiating $A_\mu(R)$,
and using (\ref{eqn27}). 
Similarly, by using (\ref{eqn27}), it
can be shown that the resulting ${\bf E}$ and ${\bf H}$
satisfy Maxwell's equations. 

In computing the commutation relations for the photon fields in
conventional theory (see Ref.~\cite{bjorken_qf}, p. 71-2), 
one cannot follow the straight
canonical path. Instead, it is necessary to modify 
the momentum space
expansion~\cite{bjorken_qf} or sum over  all three spatial 
dimensions subtracting the
longitudinal component to obtain the transverse sum~\cite{koltun}. 
For if one does
not limit the electromagnetic field to two transverse 
components, the result
is not consistent with Maxwell's equations~\cite{bjorken_qf}. 
Since the polarization
vectors for the composite photon come from 
combinations of fermion spinors,
the only choice is to sum over the polarizations 
using (\ref{eqn18}). The
result is the same as in the (modified) 
conventional theory,
\begin{equation}
\sum_{j=1}^2 \epsilon_r^j({\bf k}) \epsilon_s^j({\bf k})
= \delta_{rs} - {k_r k_s \over k^2}, 
\label{eqn36}
\end{equation}
with r,s = 1,2,3. This is the result that Kronig~\cite{kronig}
was trying to obtain with his Eq.~(17) and (19), which are not
rotationally invariant.

\subsection{Differences from Conventional Theory}
\label{sec.differ}

Unlike conventional theory, $A_\mu(R)$ is uniquely determined 
by Eq.~(\ref{eqn13}),
derived from the composite theory. The natural 
choice for the Lagrangian is,
\begin{equation}
{\cal L}= -{1 \over 4} \left( {\partial A_\mu \over \partial x_\nu}
- {\partial A_\nu \over \partial x_\mu} \right)
 \left( {\partial A_\mu \over \partial x_\nu}
- {\partial A_\nu \over \partial x_\mu} \right)
-{1 \over 2} \left( {\partial A_\mu \over \partial x_\mu}\right)^2,
 \label{eqn43}
\end{equation}
which is relativistically covariant and does 
not require the additional
constraint of gauge invariance. The second 
term vanishes without imposing
any condition, because $ (\partial A_\mu / \partial x_\mu) = 0$ 
follows directly from (\ref{eqn26}). In fact, 
$\Phi_A = -i A_4 = 0$ and
$\nabla \cdot {\bf A} = 0$,
which is like the radiation gauge. The problem 
in conventional theory
causing one to introduce two non-physical 
polarization states in order to
satisfy Lorentz invariance~\cite{veltman}, never arises 
in the composite theory. The
composite photon theory is Lorentz invariant 
without the need for gauge
invariance.

One of the most important differences between 
the composite theory and
conventional theory is that the composite 
photon creation and annihilation
operators do not obey Bose commutation relations. 
The composite photon is
a quasi-boson~\cite{perkins1999} and the expectation values 
of its commutation relations
are:
\begin{eqnarray}
\left[ \gamma_R({\bf p}^{\prime}), 
\gamma_R({\bf p}) \right] = 0, \;
\left[ \gamma_L({\bf p}^{\prime}), 
\gamma_L({\bf p}) \right] = 0, \nonumber \\
\left[ \gamma_R({\bf p}^{\prime}), 
\gamma_R^\dagger({\bf p}) \right]
= \delta( {\bf p}^{\prime} - {\bf p}) 
(1 -{\overline \Delta_{12}}({\bf p},{\bf p})),  \nonumber \\
\left[ \gamma_L({\bf p}^{\prime}), 
\gamma_L^\dagger({\bf p}) \right]
= \delta( {\bf p}^{\prime} - {\bf p}) 
(1 -{\overline \Delta_{21}}({\bf p},{\bf p})), \nonumber \\
\left[ \gamma_R({\bf p}^{\prime}),
\gamma_L({\bf p}) \right] = 0, \;
\left[ \gamma_R({\bf p}^{\prime}), 
\gamma_L^\dagger({\bf p}) \right] = 0, 
\label{eqn46}
\end{eqnarray}
where 
\begin{eqnarray}
{\overline \Delta_{12}}({\bf p},{\bf p}) = 
\sum_{\bf k} \left| F(k,{\bf n}) \right|^2 \left[  
a_1^\dagger(p\!+\!k, {\bf n})a_1(p\!+\!k, {\bf n})  
\!+\! c_1^\dagger(k, -{\bf n}) c_1(k, -{\bf n})
\right. \nonumber \\ \left.
+ c_2^\dagger(p+k, {\bf n})c_2(p+k, {\bf n})  
+ a_2^\dagger(k, -{\bf n}) a_2(k, -{\bf n}) 
  \right].
\label{eqn47}
\end{eqnarray}
The deviation from Bose statistics is caused 
by $\overline \Delta_{12}({\bf p},{\bf p})$ and 
$\overline \Delta_{21}({\bf p},{\bf p})$, 
which are functions of the neutrino creation 
and annihilation
operators. 

We can define linear polarization 
photon operators by,
\begin{eqnarray}
\xi( {\bf p}) = {1 \over \sqrt{2}} \left[ \gamma_L({\bf p}) 
+ \gamma_R({\bf p}) \right], \nonumber \\
\eta( {\bf p}) = {i \over \sqrt{2}} \left[ \gamma_L({\bf p}) 
- \gamma_R({\bf p}) \right].
\label{eqn48}
\end{eqnarray}
A particularly interesting commutation relation is,
 \begin{equation}
[\xi( {\bf p}^{\prime}),\eta^\dagger( {\bf p})]  
= {i \over 2} \delta( {\bf p}^{\prime} - {\bf p})
[\overline \Delta_{21}({\bf p},{\bf p}) 
-\overline \Delta_{12}({\bf p},{\bf p})],  
\label{eqn49}
\end{equation}
which follows from (\ref{eqn46}) and (\ref{eqn48}). 
This commutator 
is usually small as it is
zero for states with equal numbers of right and 
left handed photons.
However it {\it cannot} vanish for all states. For if it does, 
one can readily prove
that a composite photon theory is impossible 
(Pryce's theorem). The proof~\cite{pryce,berez} 
is as follows. If the commutator of (\ref{eqn49}) 
gives zero when applied to any
state vector, then all the coefficients of 
$N_{a1}({\bf k}) = a_1^\dagger({\bf k})a_1({\bf k})$,
$N_{c1}({\bf k}) = c_1^\dagger({\bf k})c_1({\bf k})$,
etc. must vanish separately. This means $F( k,{\bf n}) = 0$,
and the composite photon does not exist. 
Thus, as Pryce and Berezinkii
have proven, the composite photon cannot be a boson.

Because of these non-Bose commutation relations, ${\bf E}$  and  
${\bf H}$  of the composite
photon field do not satisfy space-like 
commutativity~\cite{perkins1965}. 
This means that we
have a non-local theory, which might be an advantage 
(see Ref.~\cite{bjorken_qf}, pp. 3-5)
because ``it is widely felt that the divergences are 
symptomatic of a chronic
disorder in the small-distance behavior of the 
[conventional] theory.''

There appears to be an inconsistence in the Standard Model
regarding the statistics of the pion and other meson. On the one
hand they are considered to be exact bosons, while on the other hand
they are considered to be composed of fermions. We would say that
the mesons are quasi-bosons.

\section{Conclusions}
\label{sec.concl}
 
The goals of this paper are to explain the historic reasons
for the rejection of the neutrino theory of light and 
to help advance composite 
photon theories. Our
achievement is very small in comparison to what is 
required for
development of a satisfactory theory. The crucial 
problem of a neutrino-antineutrino interaction and its experiment 
ramifications has not been
addressed. Nevertheless we can draw some conclusions 
from this work.

\begin{enumerate}

\item
Combining neutrino fields following a 
method similar to that of
Fermi and Yang~\cite{fermi-yang}, resulted in 
transversely-polarized particles
which have properties very similar 
to conventional photons.

\item
Pryce's theorem (stating that it is 
impossible to form composite
photons from neutrinos) contains an 
assumption which is
unsupported and probably incorrect. 
Present experimental evidence
cannot differentiate between a 
quasi-boson photon and a boson
photon.

\item
Unlike the situation with conventional 
photon theory, Lorentz
invariance can be satisfied by the composite 
photon theory without
the need for gauge invariance or introducing 
non-physical photons.

\item
While in conventional theory the polarization 
vectors of integral
spin particles are formed in a somewhat 
ad hoc manner, the
polarization vectors in the composite 
theory are combinations of
spinors resulting from the underlying 
fermion structure.

\end{enumerate}

\end{document}